\documentclass[iop]{emulateapj}
\usepackage{color}
\def\apj{ApJ}

\slugcomment{Accepted in Astrophysical Journal(ApJ)}

\shorttitle{Role of environment on nuclear activity}
\shortauthors{A. Amiri et al.}
\usepackage[latin1]{inputenc} 
\def\simlt{\mathrel{\rlap{\lower 3pt\hbox{$\sim$}}\raise 2.0pt\hbox{$<$}}}
\def\simgt{\mathrel{\rlap{\lower 3pt\hbox{$\sim$}} \raise2.0pt\hbox{$>$}}}

\begin{document}

\title{Role of environment on nuclear activity}

\author{Amirnezam Amiri $\&$ Saeed Tavasoli }
\affiliation{Physics Dept., Kharazmi University, Tehran, Iran \\ }

\author{Gianfranco De Zotti}
\affiliation{INAF-Osservatorio Astronomico di Padova,Vicolo dell'Osservatorio 5, I-35122 Padova, Italy}

\altaffiltext{}{amirnezamamiri@gmail.com}

\begin{abstract}

Motivated by the apparently conflicting results reported in the literature on
the effect  of environment on nuclear activity, we have carried out a new
analysis by comparing the fraction of galaxies hosting active galactic nuclei
(AGNs) in the most overdense regions (rich galaxy clusters) and the most
underdense ones (voids) in the local universe. Exploiting the classical BPT
diagnostics, we have extracted volume limited samples of star forming and AGN
galaxies. We find that, at variance with star-forming galaxies, AGN galaxies
have similar distributions of specific star formation rates and of galactic
ages (as indicated by the $\text{D}_n 4000$ parameter) both in clusters and
in voids. In both environments galaxies hosting AGNs are generally old, with
low star formation activity. The AGN fraction increases faster with stellar
mass in clusters than in voids, especially above $10^{10.2}\,M_\odot$. Our
results indicate that, in the local universe, the nuclear activity correlates
with stellar mass and galaxy morphology and is weakly, if at all, affected by
the local galaxy density.
\end{abstract}


\section{Introduction}\label{sec:intro}

It has long been known that, in the nearby universe, galaxy properties such as
star formation rate (SFR), morphology and  stellar mass are strongly correlated
with the surrounding galaxy density \citep{Oemler1974, Dressler1980,
Hashimoto1998, Kauffmann2004, Baldry2006}. Massive early-type galaxies in
passive evolution are preferentially found in high density environments while
the specific SFR increases towards lower density regions. This suggests that
the environment has a substantial role in driving galaxy evolution,
although the details of the physical processes involved are still poorly
understood.

We focus here on the role of environment on nuclear activity. This is relevant
to constrain processes that trigger  and control it and to investigate its
relationship to star formation. Several analyses have been carried out
considering various definitions of the local density. It is generally agreed
that the fraction of galaxies hosting active galactic nuclei (AGN) depends on
stellar mass {Pimbblet
et al 2013} and references therein, but the dependence on environment at fixed stellar mass is
debated.

\citet{Carter2001} claimed that the AGN fraction, $F_{\rm AGN}$, is insensitive
to the environment. A similar conclusion was reached by \citet{Miller2003} who
observed that while the  fraction of star-forming galaxies decreases with
density, the fraction of galaxies with an AGN remains constant from the dense
cores of galaxy clusters to the low density field. \citet{Sabater2015} argued
that the effect of the local density is minimal at fixed stellar mass and
central star formation activity. This is in agreement with the analysis of
\citet{Yang2018} who found that the sample-averaged accretion rate onto
super-massive black holes does not show any significant dependence on
overdensity or cosmic-web environment once the stellar mass of host galaxies is
controlled, and with the finding by \citet{Karhunen2014} that quasar
environments are not significantly different from those of normal galaxies with
similar luminosities.

In contrast, \citet{Kauffmann2004} reported a strong anti-correlation of
nuclear activity on local density at fixed stellar mass for powerful AGNs
($L({\rm [OIII]}) > 10^7\,L_\odot$), generally residing in massive galaxies
with significant star formation activity. The strong dependence of the
prevalence of nuclear activity with density and interactions was found after
taking into account the effect of mass. According to \citet{Constantin2008},
however, such a trend holds only for moderately
bright ($M_{\rm r} \approx -20$), and moderately massive  ($10 <
\log(M_\star/M_\odot) < 10.5$) galaxies; their data do not show any
statistically significant excess of any type of AGN in the most massive void
galaxies relative to those in relatively crowded regions (``walls'').
\citet{Deng2012} found a different environmental dependence for low and high
stellar mass galaxies. In the former $F_{\rm AGN}$ depends very little on local
density but, in the latter, it decreases with increasing density.
\citet{Lopes2017} confirmed these conclusions.

The opposite conclusion was reached  by \citet{Manzer2014} who reported a
significant increase in the AGN fraction in group environments compared with
isolated galaxies. A larger fraction of AGN  in denser environment was also
found by \citet{Argudo2018} for quenched and red galaxies.

A related issue is the role of interactions/mergers in
triggering nuclear activity. This problem has been addressed in different ways.
\citet{Schawinski2010} looked for major morphological disturbances in a sample
of spheroidal galaxies in the process of migrating from the blue cloud to the
red sequence via an AGN phase. Other studies compared the AGN fraction in
samples of close galaxy pairs with that in a control sample of galaxies with
similar masses and redshifts but no nearby companion \citep{Ellison2011,
Ellison2015, saty_ellison et al 2014, ScottKaviraj}. The general conclusion of
these investigations is that interactions/mergers substantially enhance the
specific SFR, but the effect on AGN activity depends on the AGN type. The AGN
fraction was found to be substantially higher in close galaxy pairs
\citep{Ellison2011, Ellison2015, saty_ellison et al 2014}, implying that
mergers can trigger nuclear activity and enhance the accretion rate. However,
the fraction of low excitation radio galaxies (LERGs) was found to be
independent of projected separation, implying that LERGs are not fuelled by
mergers \citep{Ellison2015}. \citet{ScottKaviraj} reported a modest decrease of
the Seyfert fraction in close pair galaxies compared to isolated galaxies,
suggesting that either mergers may not trigger AGN activity at the close pair
stage or may trigger a different AGN type. Alternatively, the lack of a clear
link between merger features and AGN activity may be due to the large time
delay between the merger-driven starburst and the peak of AGN activity,
allowing the merger features to decay \citep{Schawinski2010}. 

The apparently contradictory results on the effect  of environment on nuclear
activity have motivated our re-analysis. In this paper, we compare the extremes
of the density distribution of large scale structures: the most overdense
regions (galaxy clusters) and the most underdense ones (voids). In
Sect.~\ref{sec:sample}, we present the samples used and describe their
properties. The results of our analysis are reported in Sect.~\ref{sec:results}
while in Sect.~\ref{sec:conclusions} we summarize our main conclusions.

\begin{figure}
	\includegraphics[width=\linewidth]{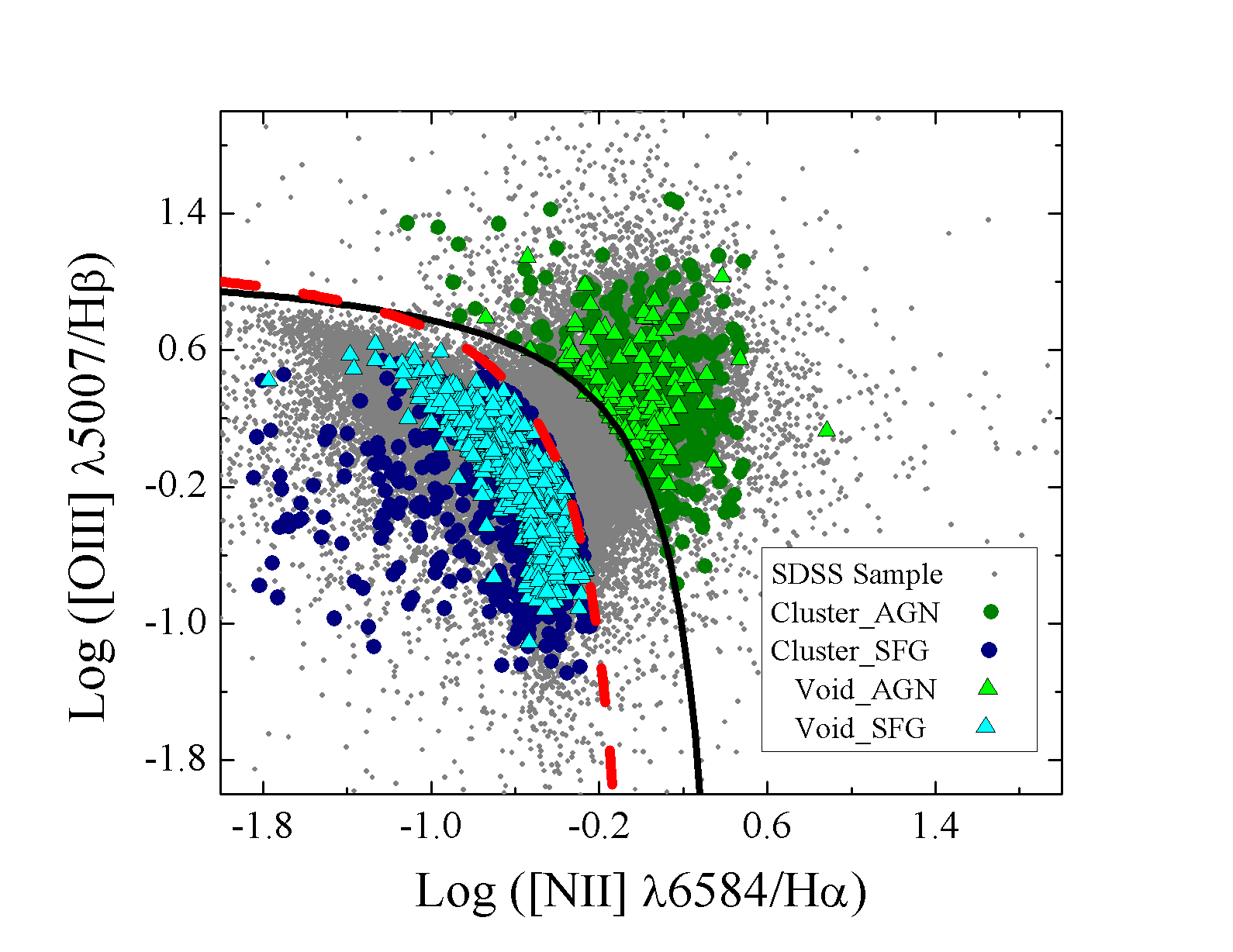}
	\caption{Diagnostic diagram to discriminate between star-forming galaxies
and galaxies hosting AGNs. The solid  black line and the dashed red line
show the separations between the two populations according to the \citet{Kewley01}
and the \citet{kauf03} criteria, respectively. Circles and triangles refer to cluster
and void galaxies, respectively. We have not classified galaxies located between the two lines.}
	\label{figure1}
\end{figure}
\vspace{2.0mm}

\section{Sample selection and galaxy classification}\label{sec:sample}

We have selected a volume-limited sample of void galaxies brighter than
$\text{M}_{r} = -18$ drawn  from the void catalogue by \citet{Tavasoli2015}.
The sample covers the redshift range $\text {0.01} < {z} < {0.04}$, the maximum
redshift being determined by the limiting absolute magnitude, and comprises 1159 galaxies.

Cluster galaxies in the same redshift and absolute magnitude range  of void
galaxies were extracted from the catalogue by \cite{Tempel14}, considering only
clusters with at least 15 spectroscopic members. Brightest cluster galaxies
(BCGs) themselves were excluded in consideration of their different evolution
compared to the other member galaxies \citep{Lauer2014}. The final sample
contains 2870 cluster galaxies.

To separate star-forming galaxies (SFGs) from galaxies hosting AGNs (hereafter
AGNs) we exploited the classical BPT diagnostics \citep{Baldwin1981} based on
emission line ratios. The emission line intensities for both void and cluster
galaxies were taken from the Max-Planck-Institute for Astrophysics (MPA)-Johns
Hopkins University (JHU) SDSS DR7 catalog \citep[MPA/JHU;][]{kauf03,
Brinchmann2004} with $\text{S/N} > 3$ for all objects. All the 1159 void galaxies and 2865 out of the 2870 cluster galaxies have emission lines and are included in the MPA/JHU catalog.

Figure~\ref{figure1} shows the distribution of our galaxies in the
[OIII]5007/H$\beta$  versus the [NII]6584/H$\alpha$ diagram together with the boundary lines between SFGs and AGNs defined by \citet{Kewley01} and by \citet{kauf03}. The
demarcation line by \citet{Kewley01} is a theoretical upper limit on the
location of SFGs in this diagram, obtained using a combination of
photo-ionization and stellar population synthesis models. It yields a
conservative AGN selection. \citet{kauf03} have revised the boundary line on
the basis of the observational evidence that SFGs and AGNs are distributed
along two separated sequences. It yields a conservative SFG selection. 

In order to avoid ambiguous classifications we have
adopted the more conservative selection for both SFGs and AGNs, excluding from
further consideration galaxies located between the two lines. The excluded
objects are a minor fraction: $\sim10\%$ of void galaxies and $\sim 16\%$ of
cluster galaxies. Thus this exclusion cannot significantly bias our results.
This classification yielded 588 cluster AGNs, 118 void AGNs, 1077 cluster SFGs
and 902 void SFGs.

To check the stability of our classification we have
re-classified our sources using the criterion by \citet{LaraLopez2010} which
exploits a different set of emission lines (H$\alpha$, [NII] and [SII]). We got
the same classification for $\simeq 97\%$ of void and $\simeq 82\%$ of cluster
SFGs, and for $\simeq 83\%$ void and $\simeq 85\%$ cluster AGNs. 

We note that while the adopted classification proved
to be solid and free from contamination by objects of uncertain type, there is
a price to pay for that: we are missing galaxies with significant contributions
from both nuclear activity and star-formation, which are located in the
intermediate region between pure SFGs and pure AGNs in the BPT diagram;
however, as already mentioned above, they are a small fraction of the initial
sample.

We are also missing strongly obscured AGNs which may
fall in the SFG region of the diagram as well as dust-enshrouded SFGs which do
not show emission lines. However only a small fraction of AGNs are heavily
obscured \citep[$7.6(+1.1,-2.1)\%$;][]{Ricci2015}, and the mean AGN fraction in
the local universe is substantially lower than that of star-forming galaxies,
as shown by the many references on that cited in Section~\ref{sec:intro}. The
fraction of dust-enshrouded SFGs is also very small \citep{HwangGeller2013}. We
can thus safely assume that both source types are too rare to significantly
affect our results.

\begin{figure}
	\includegraphics[width=\linewidth]{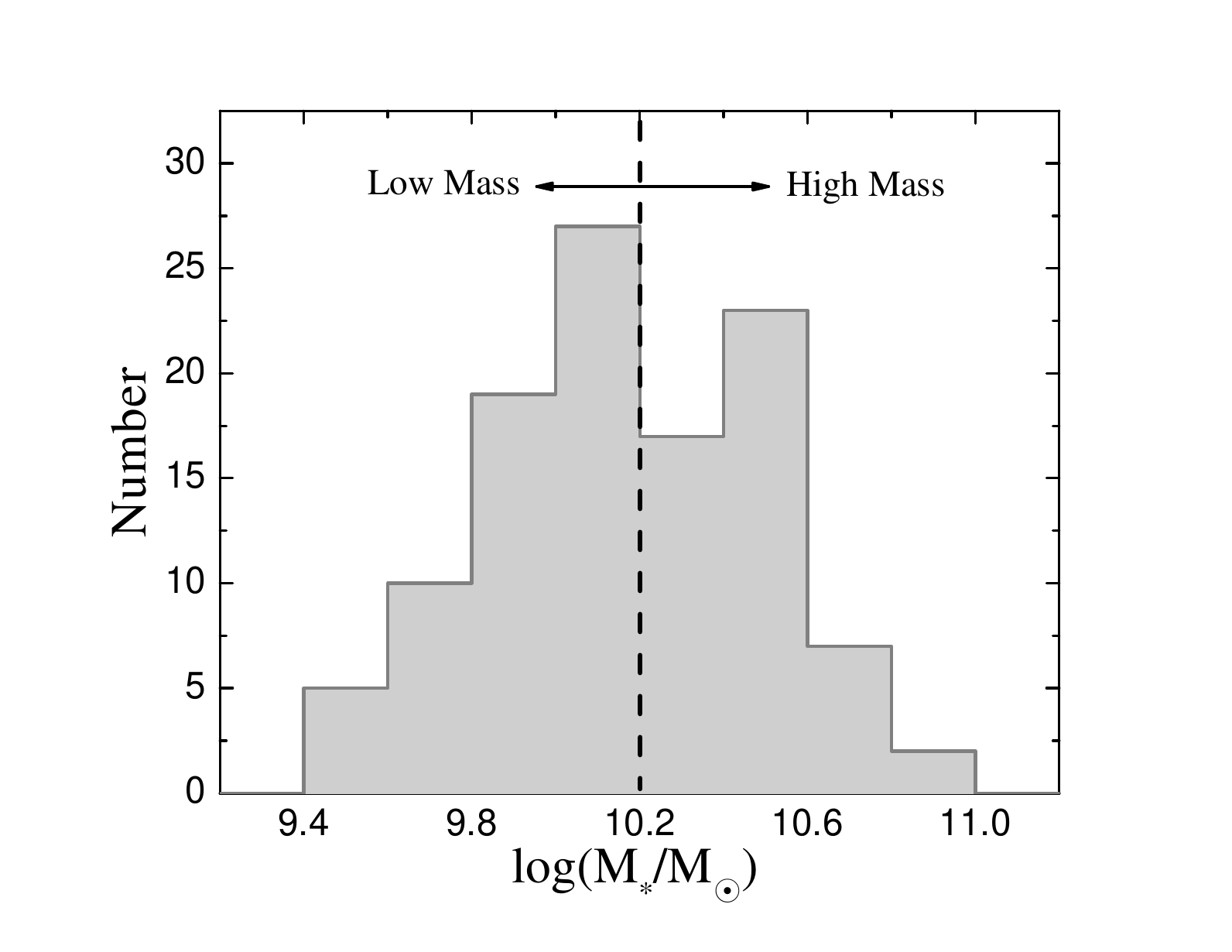}
	\caption{ Distribution of stellar masses for our
samples defined in Section\,2. The vertical dashed line shows the boundary between
low stellar mass (LSM) and high stellar mass (HSM) sub-samples. The distribution
is the same for our four source groups: SFGs (AGNs) in clusters and in voids (see text).  }

	\label{figure2}
\end{figure}

As mentioned in Section~\ref{sec:intro}, it is widely agreed that $F_{\rm AGN}$
strongly correlates with stellar mass, $M_\star$, so that the effect of
environment must be investigated at fixed $M_\star$.
To this end to compare the properties of SFGs and AGNs
in clusters and in voids in a way as homogeneous as possible, AGN and SFG
galaxies in both environments were randomly drawn from the parent samples in
such a way to have the same number of objects in each stellar mass bin so that
the stellar  mass distribution is the same for both populations
(Fig.~\ref{figure2}). Each sample has been subdivided in two stellar mass
ranges containing similar numbers of galaxies: $9.4 \leq \log(M_\star/M_\odot)
< 10.2$ (low stellar mass; LSM, 60 galaxies per subsample) and $10.2 \leq
\log(M_\star /M_\odot) \leq 11.0$ (high stellar mass; HSM, 49 objects per
subsample).
Galaxy properties come from the MPA/JHU catalog which
includes accurate aperture corrections \citep{Brinchmann2004}. For AGNs, SFRs
could not be estimated from classical SFR diagnostics like the H$\alpha$
luminosity because it is affected by the AGN component. Therefore the
$\text{D}_n 4000$ values were used instead. In turn, the star-formation
contribution affects the strengths of AGN lines, diluting their ratios. This
however does not affect our analysis since we are using line ratios only to
select AGNs and the use of the conservative \citet{Kewley01} limit ensures that
our AGN sample is clean.  

\vskip 0.75cm

\begin{figure}
	\includegraphics[width=\linewidth]{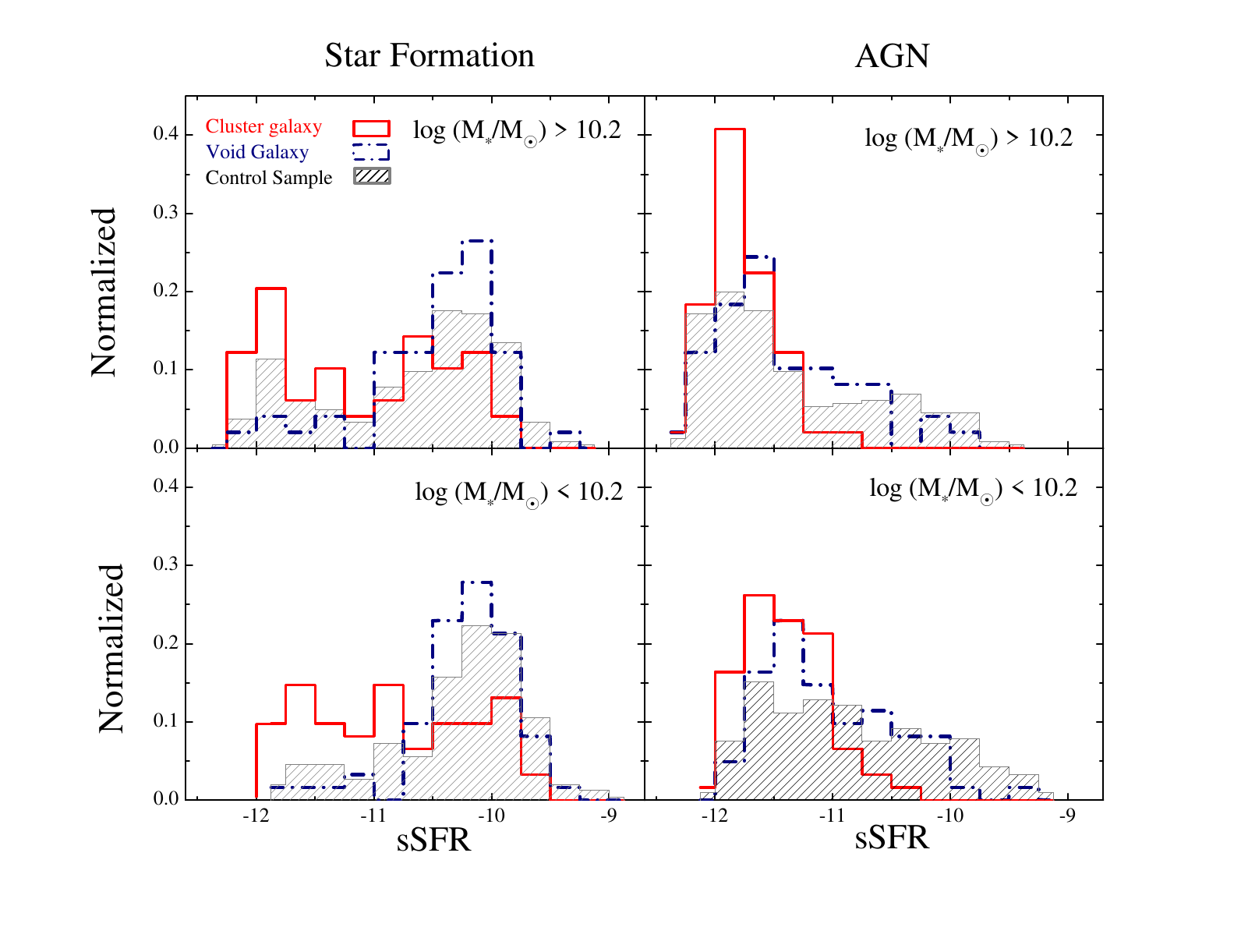}
	\includegraphics[width=\linewidth]{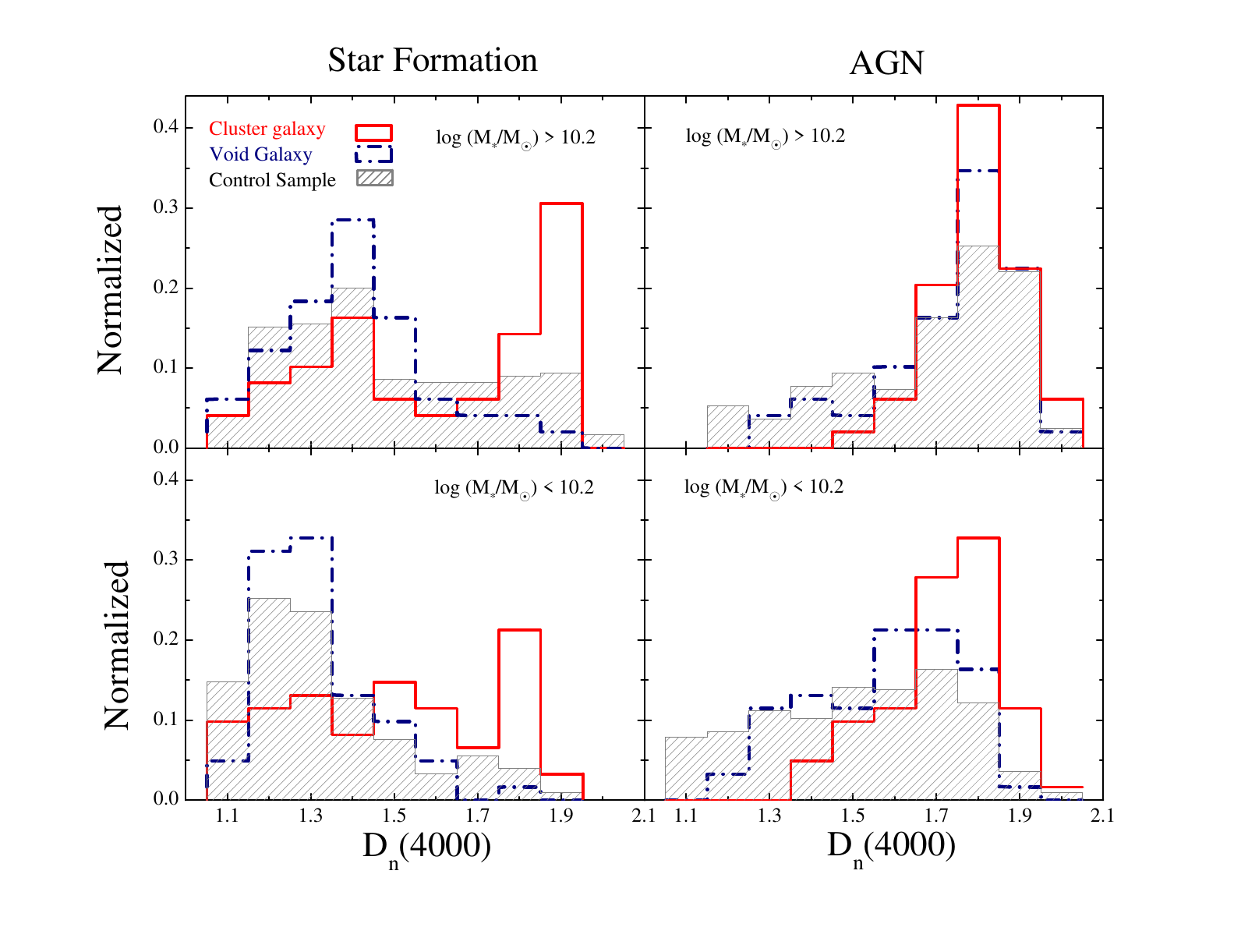}
	\caption{Distribution of specific SFRs (sSFRs; upper panel) and $\text{D}_n$4000
(lower panel) for SFGs and AGNs in clusters (red solid line) and voids (blue dash-dot line),
subdivided in two bins of stellar mass compared to the control sample (gray). }
	\label{figure3}
\end{figure}

\section{Results}\label{sec:results}

The four upper panels of Fig.~\ref{figure3} contrast
the distributions of specific SFRs (sSFRs, i.e. SFRs per unit stellar mass) in
clusters and in voids for SFGs and AGNs subdivided into LSM and HSM galaxies.
The distributions for field SFGs and AGNs in the same ranges of redshift and
stellar mass (control samples) are also shown for comparison. The control
samples where built randomly selecting from the SDSS DR7 catalog galaxies with
the same criteria used for our SFG and AGN samples, excluding those in clusters
and in voids. The selected galaxies were then classified as SFGs or AGNs in the
same way as void and cluster galaxies. In this way we obtained SFG and AGN
control samples five times larger than void and cluster samples, i.e.
containing 545 galaxies each. 
In both stellar mass ranges the distribution of sSFRs
of cluster SFGs galaxies is broad and extends to low values, consistent with
the notion that the central regions of clusters are populated by red galaxies
in essentially passive evolution, while star-forming late-type galaxies
populate the cluster outskirts. The red, passive population is rare in voids,
whose sSFR distribution is concentrated at relatively large values. The
distribution of the control sample is intermediate between those in clusters
and in voids (although somewhat closer to that in voids) for the HSM galaxies
but much closer to that in voids for LSMs. A quantitative comparison is
provided by the probabilities of the null hypothesis (same parent distribution
for different SFG samples) for void/cluster, void/control sample and
cluster/control sample SFGs given by the Kolmogorov-Smirnov (KS) test, listed
in Table~\ref{tab:KS}.
The distributions of sSFRs of void and cluster AGNs are more similar, with a
single peak at quite low sSFRs. However the
distributions for both HSM and LSM galaxies in voids have a substantial tail
extending to much higher sSFRs than in clusters. Again the distributions of
control samples are more similar to those in voids, especially for HSM
galaxies; see Table~\ref{tab:KS} for a quantitative comparison with the KS
test. 

\begin{deluxetable}{ll*{6}{c}}
\tablecaption{KS probabilities, $P$, that
distributions of sSFRs or of $\text{D}_n$4000 for different sample pairs of
SFGs and AGN galaxies are drawn from the same parent population. $P= 0.0$ means
that the probability is $<0.001$. } \tablehead{\colhead{} & \colhead{} &
\colhead{} & \multicolumn{2}{c}{\textbf{LSM}} &
\colhead{} &\multicolumn{2}{c}{\textbf{HSM}} \\
\colhead{}  &\colhead{} & \colhead{} & \colhead{sSFR} &
\colhead{$\text{D}_n$4000} & \colhead{} & \colhead{sSFR} &
\colhead{$\text{D}_n$4000}} \startdata
& Void-Cluster& & 0.0 & 0.0 & & 0.0 & 0.0  \\
\textbf{SFG}& Void-Control && 0.355 & 0.916 & & 0.089 & 0.052 \\
& Cluster-Control && 0.0 & 0.0 & & 0.016 & 0.008 \\
& & &  && &  &   \\
&Void-Cluster& & 0.002 & 0.0& & 0.002 & 0.147 \\
\textbf{AGN} &Void-Control& & 0.014 & 0.041 & &0.147 & 0.052 \\
& Cluster-Control& & 0.0 & 0.0 & &0.0 & 0.0  \\\label{tab:KS}
\end{deluxetable}

The lower panels of Fig.~\ref{figure3} carry a similar
message. The distributions of $\text{D}_n$4000, a proxy of stellar population
age, of cluster SFGs show a peak at large values, indicative of old stellar
populations. The peak is more conspicuous and is shifted to somewhat larger
values for HSMs, consistent with the ``downsizing'' scenario according to which
the more massive galaxies formed most of their stars earlier on
\citep[e.g,][]{Juneau 2005}. The distributions in voids peak at substantially
lower values of $\text{D}_n$4000, with a shift to higher values (but well below
the cluster peak) for HSM galaxies.  Like in the case of sSFRs, the
distribution for the control sample is intermediate between those in clusters
and in voids (but closer to the latter) for HSMs and very similar to that in
voids for LSMs (see also Table~\ref{tab:KS}).

\begin{figure}
	\includegraphics[width=\linewidth]{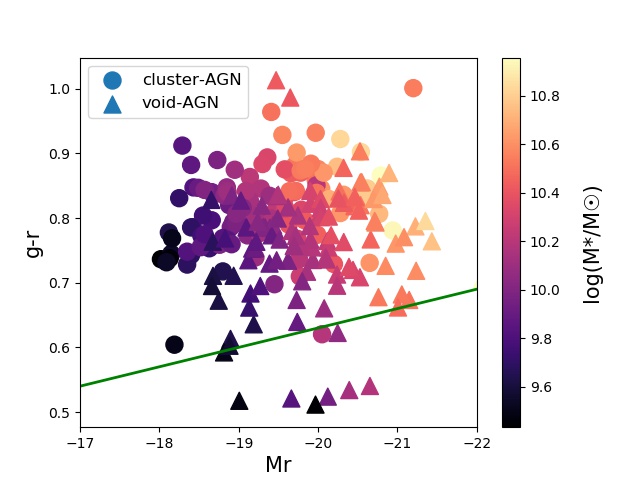}
	\caption{Color-magnitude relation for cluster (circles)
and void (triangles) AGN galaxies. The green solid line shows the luminosity dependent
threshold, defined by \citet{Bernardi et al. 2010}, which separates
early- and late-type galaxies. Almost all AGN galaxies are redder that the threshold and
have therefore early-type colors. Colors of the symbols correspond to the stellar mass
according to the scale on the right of the figure. }
	\label{figure7-1}
\end{figure}

Again the distributions are more similar in the case
of AGNs, especially for high stellar mass objects. But void galaxies have a
conspicuous tail toward low $\text{D}_n$4000's, not present in clusters but
also seen in the control sample.  Almost all AGNs in clusters and most of the HSM ones
in voids and in the field are above the boundary \citep[$\text{D}_n
4000=1.5$; e.g.]{Vergani2008} between spectroscopic late- and early-type
galaxies. An independent confirmation of this
indication is provided by galaxy colors. \citet{Bernardi et al. 2010} showed
that early-type (red) and late-type (blue) galaxies occupy different regions of
$g-r$ versus $\text{M}_r$ color magnitude plane. The boundary between red and
blue sequences is $g - r = 0.63 - 0.03 ({\rm M}_r + 20)$, shown by the solid
green line in Fig.~\ref{figure7-1}. This figure shows that almost all AGN
galaxies in clusters and in voids have early-type colors, i.e. have old stellar
populations.

\begin{figure}
	\includegraphics[width=\linewidth]{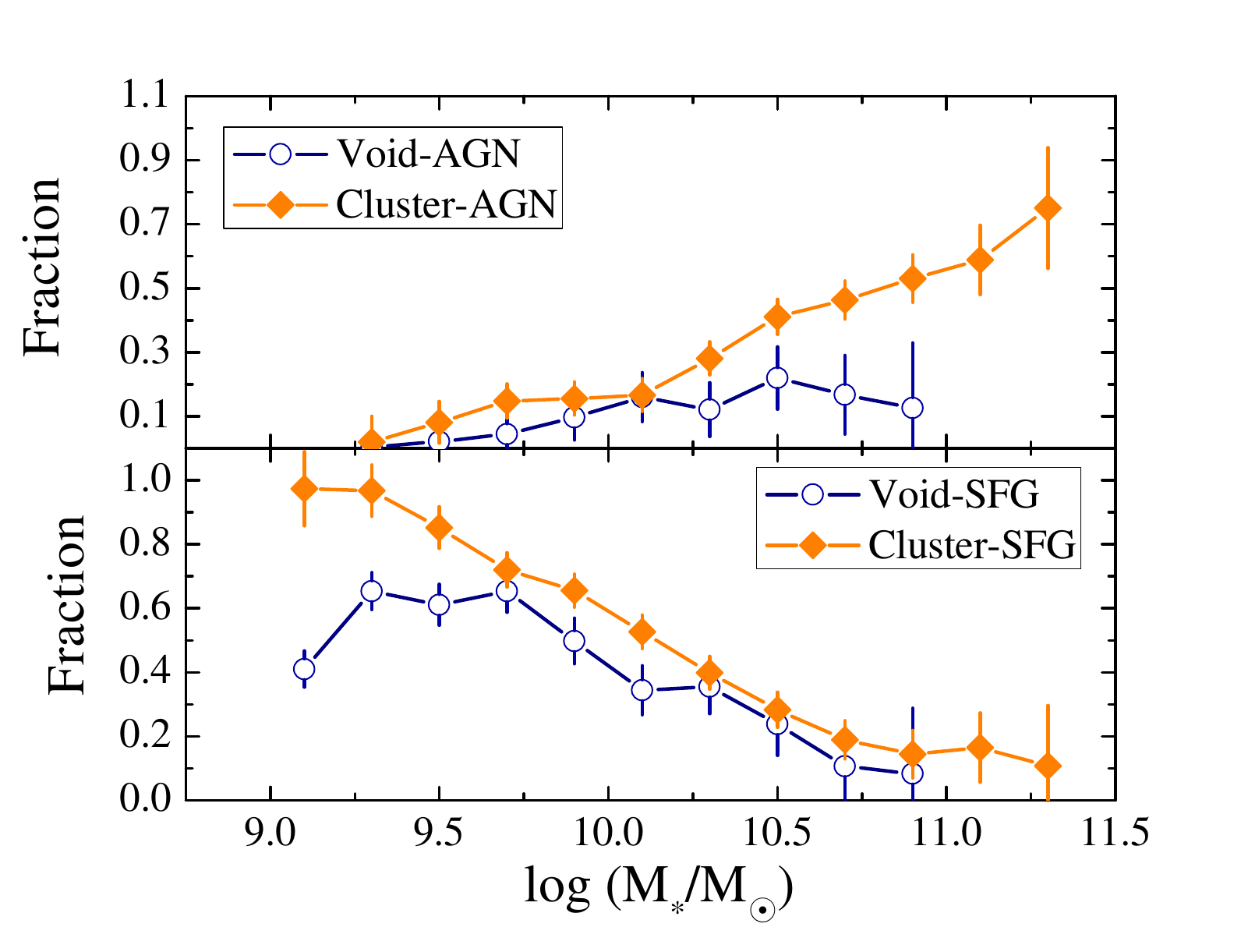}
	\caption{AGN (upper panel) and SFG (lower panel) fractions in
clusters and  voids as a function of the stellar mass.}
	\label{figure6}
\end{figure}

The main result of this paper is illustrated by Fig.~\ref{figure6} which shows
the AGN fraction, $F_{\rm AGN}$, as a function of stellar mass in clusters and
in voids. Note that for this comparison we used all cluster and void members,
not only the random selection exploited for previous analyses. At lower stellar
masses, $F_{\rm AGN}$ is only slightly but systematically higher in clusters
than in voids. Above $10^{10.2}\,M_\odot$ the mass dependence flattens out in
voids and steepens in clusters, where it reaches $\simeq 70\%$ at the highest
stellar masses.
The finding of a higher AGN fraction in dense
environments is consistent with the results by \citet{Manzer2014} and by
\citet{Argudo2018}, but disagrees with the results of many other studies (see
Sect.~\ref{sec:intro}). The reasons for the disagreements are not totally
clear, but it is likely that the different selection criteria have an important
role in that. In particular, it is clear from
Fig.~\ref{figure6} that the higher AGN fraction in clusters is due, in part, to
the higher abundance of massive galaxies that are more likely to host AGNs. The
shortage of massive galaxies is specific to voids. 
The lower panel of Fig.~\ref{figure6} shows that the fraction of SFGs follows
the opposite trend to $F_{\rm AGN}$. In both environments $F_{\rm SFG}$
increases with decreasing stellar mass down to $\simeq 10^{9.7}\,M_\odot$. At
lower stellar masses $F_{\rm SFG}$ continues to grow in clusters, reaching
100\% at the lowest masses. In the voids, somewhat surprisingly, $F_{\rm SFG}$
first flattens out and then declines at the lowest masses.
Both the increase of the AGN fraction and the decrease
of the SFG fraction with increasing stellar mass in dense environments agree
with the results by \citet{Argudo2018}.

\vspace{2.0mm}

\begin{figure}
	\includegraphics[width=\linewidth]{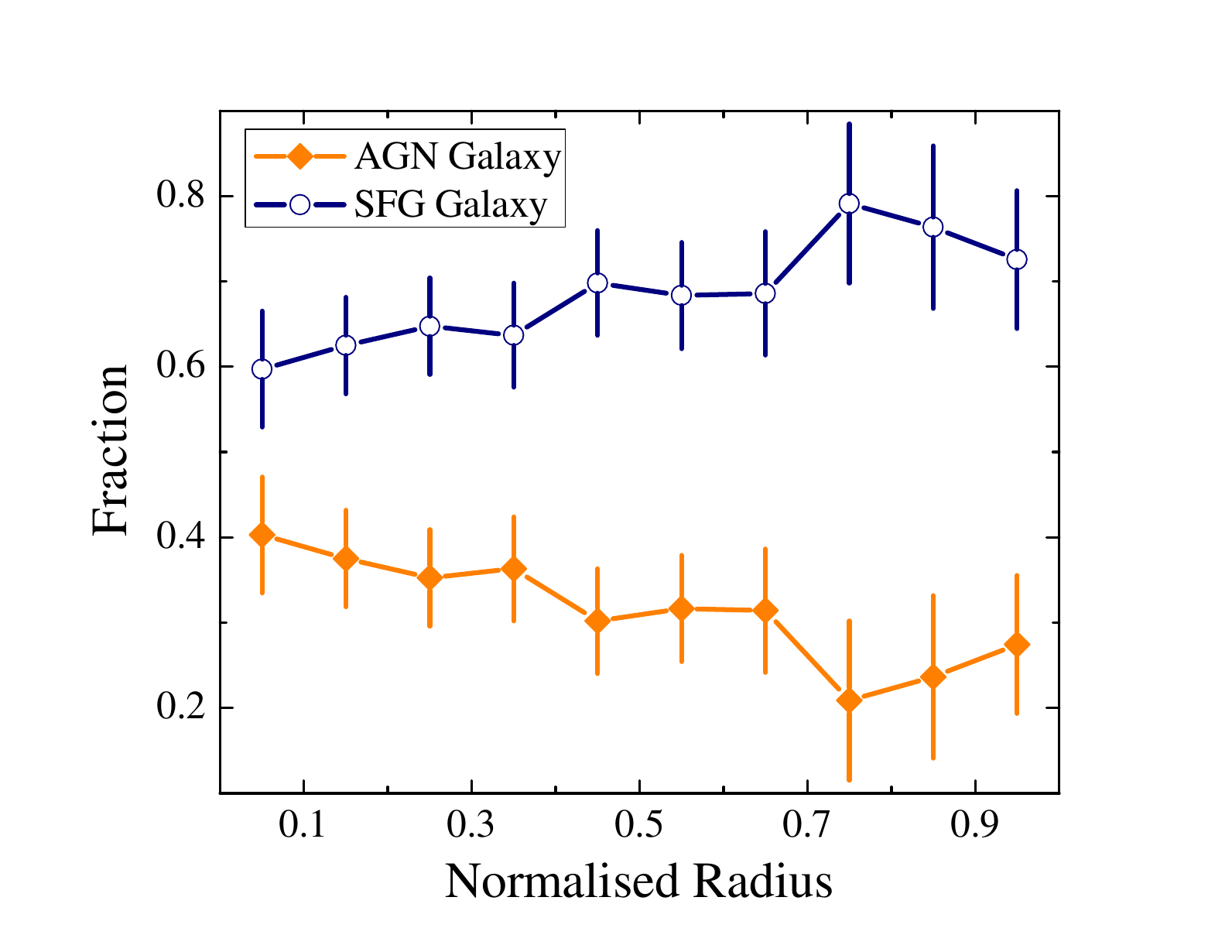}
	\caption{Average AGN and SFG fractions in clusters as a function of the distance from the brightest cluster member.}
	\label{figure7}
\end{figure}
\vspace{2.0mm}

Figure~\ref{figure7} shows the average  AGN and SFG fractions in clusters
versus the distance from the brightest cluster galaxy, assumed to be located
close to the cluster center. Again, all cluster members are used for this
analysis. The fraction of SFGs increases with
increasing distance, consistent with the notion that the sSFR is larger in the
outer regions of clusters. The AGN fraction is very weakly dependent on
distance, with a slight indication of a decline at the largest distances.
This is consistent with the evidences illustrated by Fig.~\ref{figure3} that
AGNs are preferentially associated to red galaxies. On the other hand, this
result is at odds with \citet{Lopes2017} who claimed a decline of $F_{\rm AGN}$
with decreasing distance from the cluster center. These authors, however, have
only considered massive objects, ($\log(M_\star/M_\odot) > 10.6$).

\section{Conclusions}\label{sec:conclusions}

We have addressed the still unsettled issue  of the role of environment on
nuclear activity by comparing the fraction of galaxies hosting AGNs in local
(${0.01} < {z} < {0.04}$) galaxy clusters and voids, as a function of the
stellar  mass of the host galaxies.
To this end, we have exploited volume limited galaxy samples drawn from the
void catalogue by \citet{Tavasoli2015} and from the cluster catalogue by
\cite{Tempel14}. We have extracted sub-samples of star forming galaxies and of
galaxies hosting AGNs by means of the classical BPT diagnostics. To be
conservative, the assignments to each population were done using the more
restrictive among the \citet{kauf03} and the \citet{Kewley01} criteria;
intermediate galaxies were discarded.
The distributions of both sSFRs and  $\text{D}_n 4000$ in clusters  and in
voids for SFGs differ since clusters host an essentially passive (very low
sSFRs), old ($\text{D}_n 4000\simgt 1.5$, corresponding to spectroscopic
early-type galaxies) galaxy population, missing in voids. Such distributions
are more similar to each other in the case of galaxies hosting AGNs: both in
clusters and in voids they peak at low sSFRs. The distributions of both
quantities converge in indicating that both in clusters and in voids galaxies
hosting AGNs are generally old, with low star formation activity. This is in
keeping with the well established fact that super-massive black holes correlate
with spheroidal components (bulges) of galaxies, not with disks
\citep[e.g.,][]{KormendyHo2013}.
The AGN fraction increases with stellar mass both in clusters and in voids. The
mass dependence is, however, quite different: it is mild in voids where $F_{\rm
AGN}$ reaches $\simeq 30\%$ at most; in clusters it is steeper, particularly
above $M_\star \simeq 10^{10.1}\,M_\odot$; $F_{\rm AGN}$ reaches $\simeq 70\%$
at the highest masses. The higher $F_{\rm AGN}$ in clusters, compared to voids,
agrees with the results by \citet{Manzer2014} and by \citet{Argudo2018}, but
disagrees with the results of many other studies which however have used
different selection criteria.
The average SFG fraction in clusters is higher in the
less dense outer regions, where the fraction of star-forming galaxies is lower.
This is consistent with the well known relation between density and
morphological type in the local universe \citep[e.g.,][]{Dressler1980}:
star-forming galaxies prefer low-density environments. 
The AGN fraction is found to be very weakly dependent
on the distance from the cluster center, with a hint of a decrease at the
cluster boundary. This finding is at odds with \citet{Lopes2017} who claimed a
decline of $F_{\rm AGN}$ with decreasing distance from the cluster center.
Our results indicate that, both in local clusters and
in local voids, the nuclear activity is related to galaxy properties such as
stellar mass and morphology and is only weakly, if at all, affected by
environment.  We caution, however, that this conclusion
may not apply to intermediate-density systems. For example, it is possible that
the extreme environments we have investigated have in common a low
merger/interaction rate. In voids close encounters are rare due to the low
galaxy density; in clusters  merging is ineffective when galaxy velocities are
much larger than their stellar velocity dispersions
\citep[e.g.,][]{Carnevali1981}. The exception are BCGs that cannibalized many
satellite galaxies during their lifetime; they are also excluded from our
samples.
\acknowledgments
We thank the anonymous referee for a very thorough
review that helped us to significantly improve the quality of this paper and R.
van de Weygaert for a fruitful discussion on galaxy evolution in voids. This
paper has made use of SDSS data. Funding for the SDSS and SDSS-II has been
provided by the Alfred P. Sloan Foundation, the Participating Institutions, the
National Science Foundation, the U.S. Department of Energy, the National
Aeronautics and Space Administration, the Japanese Monbukagakusho, the Max
Planck Society, and the Higher Education Funding Council for England. The SDSS
Web site is http://www.sdss.org/. The SDSS is managed by the Astrophysical
Research Consortium for the Participating Institutions. The Participating
Institutions are the American Museum of Natural History, Astrophysical
Institute Potsdam, University of Basel, University of Cambridge, Case Western
Reserve University, University of Chicago, Drexel University,Fermilab, the
Institute for Advanced Study, the Japan Participation Group, Johns Hopkins
University, the Joint Institute for Nuclear Astrophysics, the Kavli Institute
for Particle Astrophysics and Cosmology, the Korean Scientist Group, the
Chinese Academy of Sciences (LAMOST), Los Alamos National Laboratory, the
Max-Planck-Institute for Astronomy (MPIA), the MaxPlanck-Institute for
Astrophysics (MPA), New Mexico State University, Ohio State University,
University of Pittsburgh, University of Portsmouth, Princeton University, the
United States Naval Observatory, and the University of Washington.

\clearpage

\end{document}